\documentclass[12pt]{iopart}

\usepackage{epsf}
\usepackage{graphicx}
\usepackage{color}
\definecolor{red}{rgb}{1.,0.,0.}

\begin{document}

\title[Inferring basic     parameters of the geodynamo from sequences of polarity reversals]
{Inferring basic parameters of the geodynamo from sequences of polarity reversals}

\author{M Fischer, G Gerbeth,  A Giesecke and F Stefani}

\address{Forschungszentrum Dresden-Rossendorf, P.O. Box 510119,
D-01314 Dresden, Germany}
\ead{F.Stefani@fzd.de}
\begin{abstract}
The asymmetric time dependence
and various statistical properties of 
polarity reversals of the Earth's magnetic field are utilized
to infer some of the most essential parameters of the geodynamo,
among them the effective (turbulent) magnetic diffusivity,
the degree of supercriticality, and 
the relative strength of the periodic forcing
which is believed to result from the Milankovic cycle of the 
Earth's orbit eccentricity.
A time-stepped spherically symmetric $\alpha^2$-dynamo model 
is used as the kernel of an inverse problem solver in form of a
downhill simplex method which 
converges to solutions that yield a stunning
correspondence with paleomagnetic data.

\end{abstract}

\submitto{\IP}
\maketitle

\section{Introduction}

The hydromagnetic dynamo in the Earth's outer core
converts gravitational and thermal energy into magnetic
energy \cite{MERRILL}.
One of the most impressive feature of the geomagnetic field
is the irregular occurrence of polarity reversals.
Averaged over the last few million years the
mean rate of reversals is approximately 4-5 per Myr, 
although the last reversal occurred approximately
780000 years ago.
At least two, but very likely  three \cite{GALLET,COURTILLOT} 
superchrons have been identified
as ''quiet'' periods of some tens of millions of years showing
no reversal at all. 

Knowledge on ancient magnetic field data is mainly obtained from paleomagnetic measurements
of permanent magnetization from (frozen) lava and sedimentary rocks. However,
appropriate paleomagnetic sites are rare, unevenly distributed across
the Earth's surface and, furthermore,  actual dating methods allow only a rather rough
time-resolution. 
Hence only few reversal characteristics can be evaluated as robust \cite{MERRILL2}.
One of the commonly accepted   
features of reversals is
their pronounced temporal asymmetry  with the initial
decay of the dipole being
much slower than the subsequent
recreation of the dipole with opposite
polarity \cite{VALET}.

Recent numerical simulations have been successful in
reproducing not only the dominance of the axial dipole and
the spectrum of the geomagnetic field, but also the irregular
occurrence of polarity reversals \cite{ROGLA,WICHTOLSON,AUBERT}. 
Polarity reversals were also
observed in one \cite{BERHANU} of the recent liquid sodium 
dynamo experiments which have flourished during 
the last decade \cite{RMP}.

It is important to note, however, that neither in simulations 
nor in experiments it is
possible to accommodate all dimensionless parameters of the geodynamo 
\cite{GUBBINS,ZAMM}, and many of them are not even well known 
\cite{ROGLA2}.

A complementary way to acquire knowledge about the 
geodynamo is 
to use available magnetic field data for constraining the
source of dynamo action. Most famous 
among those attempts is the frozen-flux approximation 
\cite{ROSCO,HOLMEOLSEN} 
which allows to infer the tangential flow at the 
core-mantle-boundary from secular variation of the 
geomagnetic field. Going beyond this frozen-flux approximation
and trying to infer properties of the geodynamo in the
depth of the Earth's core has been less successful so
far. Such trials to  ''look inside the dynamo'' by utilizing
spectral data  worked nicely for simplified
dynamo models \cite{AN,PEPI,PRE} but are  hardly  
applicable for real world dynamos.

With this sobering experience in mind, in the present paper we 
undertake a rather uncommon attempt to
constrain some of the most significant parameters of the 
geodynamo by various characteristics of paleomagnetic 
reversal records.
Most prominent among those characteristics is the 
above mentioned temporal asymmetry  of reversals.
Two further characteristics reflect some sort of ordering
in the otherwise irregular reversal sequence.
The first one is the clustering property of
reversals which was discovered only recently \cite{CARBONE,LUCA}. 
Clustering of reversals is manifested in an enhanced probability 
of a
consecutive reversal shortly after 
a first reversal has occurred,  which results in a deviation 
from the
Poisson distribution that would hold for uncorrelated events.
The second one is
the appearance of a $\sim$100 kyr periodicity in the
distribution of the residence times (RTD)
between reversals which is believed to result 
from the Milankovic cycle of the 
Earth's orbit eccentricity 
\cite{CONSOLINI,LORITO}.

Based on these three input features we will 
examine in this paper 
a simplifying $\alpha^2$-model of the geodynamo for which we
estimate the degree of supercriticality, the noise level, the relative
strength of the periodic forcing, 
and the effective (turbulent) magnetic diffusivity of the Earth's
outer core. Actually, a few dependencies on individual parameters 
were already published in preceeding papers. In \cite{EPSL} the 
dependence of typical time scales on the supercriticality of the 
dynamo was
studied, in \cite{LUCA} some dependencies of the 
clustering property
on the supercriticality and the
noise level were shown, and in \cite{EPJB} 
the influence of the diffusion time scale 
on the RTD between reversals was touched upon.
What is new in the present paper is that we 
take all three reversal features together 
and  try to infer from 
them some essential parameters of the geodynamo.

We will start with a presentation of the forward 
dynamo problem for which we will use a rather simple, 
spherically symmetric
mean-field dynamo of the $\alpha^2$ type.
This simple model had turned out to be  quite 
helpful for understanding 
the basic principle of the reversal process
as a
noise-induced relaxation-oscillation in the vicinity of an
exceptional point of the spectrum of the
non-selfadjoint dynamo operator 
\cite{EPSL,PRL,GAFD}.
This exceptional point, at which two real
eigenvalues
coalesce and continue as a complex conjugated pair of
eigenvalues,  is associated with
a nearby local maximum of the
growth rate situated at a slightly lower
magnetic Reynolds number.
It is the negative slope of the growth rate
curve between this local
maximum and the exceptional point that makes
stationary
dynamos vulnerable to  noise.
Then, the instantaneous
eigenvalue is driven
towards the exceptional point and beyond into the
oscillatory branch where the sign change of the dipole polarity happens.

After having delineated the
simplified mean-field dynamo model
we will present the solution method for the inverse problem and
the main results.

The paper concludes with a summary and 
a speculation on the possible consequences of our findings
for the general understanding and the numerical 
simulations of the geodynamo.

\section{The forward problem}

Before tackling the inverse problem by evaluating 
the reversal characteristics of many different solutions 
of the time evolution equation for the magnetic field we 
have to delineate in the present section the forward problem, 
although this has been described already in \cite{EPJB}.

The governing equation for the mean magnetic field  ${\bf{B}}$ is the 
induction equation without any mean flows (${\bf{v}}=0$)
under the influence of a helical turbulence parameter $\alpha$  \cite{KRRA}:
\begin{equation}
\frac{\partial {\bf{B}}}{\partial \tau} = {\bf \nabla} \times (\alpha {\bf{B}}) +
\frac{1}{\mu_0 \sigma} {\bf \nabla}^2 {\bf{B}} \; .
\end{equation}
This equation results from pre-Maxwell's equations
when the source of the magnetic field generation is supposed not 
to be a large scale velocity $\bf v$ but some turbulence which
comprises helical parts.
Since the magnetic field is divergence-free 
we can decompose it into a poloidal and a toroidal parts according to
\begin{equation}
{\bf{B}}=-\nabla \times ({\bf{r}} \times \nabla S)-{\bf{r}} \times \nabla T \; .
\end{equation}
In spherical geometry, the two defining scalars $S$ and $T$ are easily
expanded in
spherical harmonics of degree $l$ and order $m$. In the following 
we will
assume $\alpha$ to be spherically symmetric, being well aware of the fact 
that this
grave simplification does not apply to the Earth's
outer core. The great advantage of this simplification is
that the induction equation decouples into pairs of
partial differential equations for each degree
$l$ and order $m$,
\begin{eqnarray}
\frac{\partial s_l}{\partial \tau}&=& \frac{1}{r}\frac{\partial^2}
{\partial r^2}(r s_l)-\frac{l(l+1)}{r^2} s_l +\alpha(r,\tau) t_l \; ,\\
\frac{\partial t_l}{\partial \tau}&=&\frac{1}{r}\frac{\partial}{\partial r}
\left[ \frac{\partial}{\partial r}(r t_l)-\alpha(r,\tau)
\frac{\partial }{\partial r}(r s_l) \right]- \frac{l(l+1)}{r^2}
[t_l-\alpha(r,\tau) s_l] \; .
\end{eqnarray}
where we have already used dimensionless 
parameters in which the radius $r$
is measured in units of the radius of the Earth's outer core, $R$, the time $\tau$ in units of the
diffusion time $T_d:=\mu_0 \sigma R^2$, and $\alpha$ in units of
$(\mu_0 \sigma R)^{-1}$.
The  boundary conditions are: $\partial s_l/\partial r             
|_{r=1}+{(l+1)} s_l(1)=t_l(1)=0$.                                  
In the following we will consider only the                         
dipole field with  $l = 1$.                                        

Due to the presupposed spherical symmetry of $\alpha$ there is no 
preferred direction of the magnetic field axis, hence the 
order $m$ of the spherical harmonics
does not show up in the 
equation system (3-4).
This absence of a preferred direction of the dominant dipole          
could be considered a significant weakness of our model.              
Strictly speaking, our restriction to the
axial dipole (the mode with $l=1$, $m=0$) can only be justified
if some  additional                                          
symmetry breaking mechanism is supposed to  work.
This is simply the prize we have to pay for the great advantage       
of remaining in the framework of only two coupled partial           
differential equations (3,4) for $s_1$ and $t_1$. 
Certainly, a study of more realistic dynamo models would be 
highly desirable, but with the present 
computer resources it will be difficult to get 
the good statistical validity that is easily obtainable 
with the large number of reversals resulting 
from our simple model, let alone the solution of an inverse problem 
as it will be presented in this paper.

The equation system (3,4), with fixed $\alpha(r)$,
would represent a so-called kinematic mean-field dynamo model. 
Below a critical
amplitude of $\alpha(r)$, the magnetic eigenfield would
decay exponentially, above this value it would
grow exponentially.
In reality, of course, the exponential growth of the 
magnetic field cannot continue indefinitely.
After having grown to a certain amplitude, the magnetic field
attenuates the
source of its own generation (Lenz's rule). While the precise 
way of this attenuation is an interesting 
topic in its own right, we will restrict ourselves to a 
very simple algebraic ''quenching'' of the kinematic $\alpha$ 
with the angle averaged
magnetic field energy which can be expressed in terms of $s(r)$
and $t(r)$. Note again, that this averaging over the spherical 
angle is 
another simplification that is chosen in order to remain 
in the framework of
a spherically symmetric model for which the $l$ and $m$ 
decoupling of the equation (3,4) remains valid, at least
formally.

While the non-linear system of equations that results from (3,4) and
the algebraic quenching  already exhibits a very rich behaviour
we will additionally consider the influence of noise by
which $\alpha(r)$ is influenced. 
This noise might be
considered as a shorthand for fluctuations of the flow, 
changing boundary conditions, and 
the neglected influence of higher magnetic field modes.

Summarizing the quenching and the noise effect, we model
the time dependent $\alpha(r,\tau)$ in the form
\begin{eqnarray}
\alpha(r,\tau)&=&\frac{\alpha_{kin}(r)}{1 + E \left[ {\frac{2 s_{1}^{2}(r,\tau)}{r^2}+
\frac{1}{r^2}\left( \frac{\partial (r s_{1}(r,\tau))}
{\partial r} \right)^2 + t_{1}^{2}(r,\tau) } \right]   } \nonumber \\
&&+  \xi_1(\tau) + \xi_2(\tau)
 r^2 + \xi_3(\tau)  r^3+\xi_4(\tau) r^4 \; ,
\label{alpha}
\end{eqnarray}
In this equation the noise is considered to have 
a finite correlation time in which it is supposed to be constant. 
This is equivalent to the following  temporal correlation:
\begin{eqnarray}
< \xi_i(\tau) \xi_j(\tau+\tau_1)>&=&D^2 (1-|\tau_1|/T_{c}) \times \Theta(1-|\tau_1|/T_{c}) \delta_{ij} \; ,
\end{eqnarray}
where $\Theta$ is the Heaviside step function.
In equations (5,6), $\alpha_{kin}(r)$ is the kinematic $\alpha$ profile,
$D$ is
the noise intensity, $E$ is a
constant measuring the inverse mean magnetic field energy,
and $T_c$ is
a correlation time of the noise. 

In the following we will motivate the particular choice of the
$\alpha(r)$ profile. 
In former 
papers it was shown that kinematic dynamos of oscillatory character
appear only in a rather small corridor of $\alpha(r)$ profiles
that are characterized by at least one sign change along the radius.
While this was first shown for the spherically symmetric
$\alpha^2$ dynamo in \cite{PRE}, a quite similar result was
later obtained for a more realistic model in which the 
latitudinal dependence of $\alpha$ is governed by a $\cos{\theta}$
dependence \cite{GIESECKE}. Interestingly, such $\alpha$ profiles 
with one sign change along $r$ 
were indeed found in simulations of magnetoconvection in the
Earth's outer core \cite{GIESECKE2}.

Based on this motivation, we choose for
the kinematic $\alpha$ profile in equation (5)
the particular Taylor expansion                                       
\begin{eqnarray}
\alpha_{kin}(r)=C  \cdot ( \alpha_0 + \alpha_{1} r
+ \alpha_{2} r^{2} + \alpha_{3} r^{3} +
\alpha_{4} r^{4})
\end{eqnarray}
with 
$\alpha_1=\alpha_3=0$, $\alpha_2=-6$ and $\alpha_4=5$. 
The first coefficient, $\alpha_0$, is chosen close to 1, but with
two important modifications. First, we add a small parameter $\delta$ 
which regulates the proximity of the kinematic dynamo to oscillatory solutions.
Second, $\alpha_0$ will also incorporate the periodic forcing with the
dominant 95 kyr period of the Milankovic cycle,
being well aware of the fact that the
Milankovic cycles contain also other frequencies \cite{LIU}.
Hence, taking both effects together we end up with the following 
{\it ansatz} for the time-dependence of $\alpha_0$:
\begin{eqnarray}
\alpha_0(\tau) = 1 + \delta+\epsilon \cos( \frac{2 \pi}{T_{\Omega}} \cdot \tau),
\end{eqnarray}
where $\epsilon$ parameterizes the strength of the periodic forcing.
One could ask why only assuming the first Taylor expansion coefficient
as time-dependent and not the entire expression on the r.h.s. 
of equation (7). The reason for this is that such a homogeneous 
scaling would have only a weak influence on the dynamo, since
it results simply in a stronger quenching of $\alpha_{kin}(r)$
ending up (approximately)
at the same quenched $\alpha(r)$ profile as before. In contrast to this,
a selective change of only one Taylor expansion coefficient in
equation (7) changes the {\it shape} of the $\alpha(r)$ profile,
which is much more effective for  changing the reversal probability.
Going back to the real geodynamo, it is also not very likely 
that the Milankovic cycle of the Earth's orbit eccentricity 
changes the dynamo source homogeneously.

The equation system (3-5), with the concretization  (6-8),
is time-stepped by means of a standard
Adams-Bashforth method with radial grid spacing of 0.02 and
time step length of 2 $\times 10^{-5}$. The correlation time $T_c$
has always been
set to $ 0.005 \cdot T_c$ which would correspond
to 1 kyr in case that the
diffusion
time is set to 200 kyr.
The resulting time series show reversal sequences
quite similar to those of the geodynamo
\cite{EPSL,GAFD,MAHYD}. 
For the sake of simplicity, we define a reversal as the               
sign change of the poloidal field component $s(1,\tau)$               
at the  outer radius $r=1$. Depending on the precise parameters, we
get typically some 10$^4$ reversals in 1 day CPU time on a normal workstation. 
As an important characteristic
of these sequences we will determine the
distribution of
residence times (RTD) $\tau_r$ between two subsequent reversals.

\section{The inverse problem}

In this section we will try to determine  some of 
the parameters of the presented
dynamo model in such a way that the resulting 
reversal sequences  fit the characteristics of 
paleomagnetic data as accurate as possible.

\subsection{The parameters to be determined}

Being well aware of some arbitrariness in the 
particular choice of free parameters in the dynamo model, 
we have decided to choose five parameters which we 
believe to represent some generic features of every 
dynamo model.

The first one, with a comparably clear 
relevance and interpretation, 
is the magnetic Reynolds number $C$ (that is based, however, 
not on the velocity but on the helical
turbulence parameter $\alpha$). The importance of this parameter
comes from the fact that, as a measure of the 
supercriticality of the
dynamo, it governs the typical time scale
of the reversal process. This can be understood as follows:
While in the rather ''quiet'' regime, when the dynamo
is in one polarity, the value of $\alpha$ 
is quenched approximately to the critical one, amidst
a reversal (when the magnetic field energy becomes small)
$\alpha(r)$ will get close to the unquenched (kinematic) profile.
This means that during this short time interval
the magnetic field 
dynamics is dominated by the possibly very large 
instantaneous growth rates (and frequencies) of the
kinematic $\alpha(r)$ profile.
Roughly speaking, the higher $C$ the faster the reversal process 
happens (see, e.g., Fig. 8 in \cite{EPSL}). 
In addition to this, $C$ governs also the asymmetry of reversals. 
To  understand this effect
imagine a kinematic $\alpha(r)$ profile which gives an oscillatory dynamo
at the critical value of $C$. In this case, ''reversals'' are 
nothing than parts of harmonic oscillations.  With increasing $C$, i.e. with increasing supercriticality, these 
oscillations become more and
more anharmonic (relaxation oscillations) \cite{GAFD,PRL}. 
Very often, then, there is another transition value of $C$ 
beyond which 
the (noise-free) oscillatory dynamo becomes stationary again. 
However, even 
in this case reversals can be triggered by noise, and the 
asymmetry will still be  governed by $C$.

The second parameter to be determined 
is the noise level, represented by
$D$ in equation (6). Roughly speaking, the larger $D$, 
the more frequent the system will leave the stable 
regime and undergo a reversal process. A quantitative 
interpretation of $D$ is non-trivial, 
since for our case of finite correlation times 
the relevant quantity is always $D/\sqrt{T_c}$, so any 
interpretation of $D$ is only
sensible in combination with the correlation time of the noise.

The interpretation of the third parameter $\delta$ in equation (8)
is perhaps the most intricate one.
Formally, $\delta$ describes
a constant shift of the kinematic $\alpha$ profile. 
In the purely kinematic regime, the proximity of the dynamo to an 
oscillatory solution  is very sensitive on this parameter. In
the highly supercritical regime, this sensitivity is reduced 
but there is still some influence of $\delta$ on the
probability that a reversal occurs.

The fourth and the fifth parameter refer to the interplay 
between magnetic diffusion  and the periodic forcing due to 
Milankovic cycle of the Earth's orbit eccentricity.
This problem was already addressed in \cite{EPJB} where we had 
found that the appearance of several clear-cut maxima in the RTD 
at multiples of 95 kyr is hardly compatible with a diffusion time 
$T_d:=\mu_0 \sigma R^2=227$ kyr. The latter would follow
from recent estimates \cite{STACEY} of the molecular conductivity 
($\sigma=4.71 \times 10^5$ ($\Omega$ m)$^{-1}$) of the
Fe-Ni-Si alloy
at the tremendous pressure
of the Earth's core. Such a large diffusion time 
would simply ''smear out'' the various maxima in the RTD which 
are believed to result from a stochastic resonance effect.
The interesting point is now
that  this molecular value of the conductivity 
could possibly be reduced by
the turbulent motion of the fluid.
This so-called $\beta$ effect has been 
estimated theoretically in various limiting cases \cite{KRRA}.
There have also been claims \cite{REIGHARD} on the measurement
of a few percent $\beta$ effect in a turbulent
liquid sodium flow with magnetic Reynolds numbers $Rm$
up to 8, and some indications of it have been
found in the Perm dynamo experiment \cite{PERM} and the Madison 
dynamo experiment \cite{SPENCE}, but a
clear experimental demonstration of it is still elusive.
For the geodynamo with its large magnetic 
Reynolds number and high turbulence level, a 
reduction of $\sigma$ by a factor 2 or so is not out 
of range.
This is the reason why we will treat the diffusion time 
of the dynamo 
model as the fourth free parameter to be determined. 

At the same 
time we will keep the 
period of the forcing fixed to 95 kyr (although the
actual forcing contains much more frequencies \cite{LIU}).
As indicated already in \cite{EPJB} we expect that the solution of 
the inverse problem 
will give us a diffusion time comparable or slightly smaller
than 95 kyr, since otherwise the 
appearance of several maxima in the RTD is hardly explainable.

It is quite natural to consider the strength of this 
periodic forcing, $\epsilon$ in equation (8), as the 
fifth free parameter.

\subsection{The functionals to be minimized and the 
inversion method}

Since our dynamo model is quite simple, it cannot be expected 
to reproduce all possible reversal features. In particular, 
spatial distributions of the field during reversals, e.g.  
preferred paths of the virtual geomagnetic pole (VGP), cannot
be addressed. However, we believe that 
typical temporal features of reversals,
including the shape of individual reversals, their clustering 
properties and the distribution of inter-reversal times, should
still be reproducible by the model.
These three generic characteristics seem to us 
only dependent on the general reversal
mechanism and on some basic parameters of the geodynamo, five 
of which were discussed above.

The first functional to be minimized results from the temporal 
shape of individual reversals which we have adopted from figure 4
of \cite{VALET}. Actually, the curve ''Real'' in our 
figure 1 shows the time evolution 
of the virtual axial dipole moment (VADM) averaged over the last
five reversals, between -53 kyr and 17 kyr.
The first functional $F_{shape}$ reflects the deviation of the 
simulated curves from this
''real'' curve:
\begin{eqnarray}
F_{shape}=\sum_{i=-53}^{17} ({\mbox{VADM}}_{real}(i \Delta t)-
{\mbox{VADM}}_{num}(i \Delta t))^2
\end{eqnarray}
Here, the subscript $real$ refers to the averaged paleomagnetic
measurement, and $num$ to the numerically obtained ones.
Note that for the latter we have taken an average 
over 100 reversal events, since
we thought that an average over only five reversals would result in
artificial variations with no physical significance.
$\Delta t$ has been set to 1 kyr, in accordance with the 
resolution in \cite{VALET}.

With the second functional we intend to accommodate
the clustering property of reversal events 
which had been identified by Carbone et al. in 2005 \cite{CARBONE}, 
and analyzed in more detail
in a follow-up paper by Sorriso-Valvo et al. \cite{LUCA}. 
The authors had defined a
quantity which is able to detect clusters or voids even 
for Poisson processes with time dependent rate parameters.
Consider the $i$th event in the sequence of reversals, and
denote the shorter of the preceding or following inter-reversal 
intervals (or residence times) by $\delta t$, i.e.
\begin{eqnarray}
\delta t&=&\min\{t_{i+1}-t_i; t_i-t_{i-1}).
\end{eqnarray}
Then define 
the quantity $h(\delta t, \Delta t)$ by 
\begin{eqnarray}
h(\delta t, \Delta t)&=&2 \delta t/(2 \delta t+\Delta t)
\end{eqnarray}
wherein
\begin{eqnarray}
\Delta t=t_{i+2}-t_{i+1} \;\; \mbox{if} \;\; \delta t=t_{i+1}-t_i   \;\;\;\; ,\\
\Delta t=t_{i-1}-t_{i-2} \;\; \mbox{if} \;\; \delta t=t_{i}-t_{i-1} \;\; .
\end{eqnarray}

One can easily imagine that in the case of clustering it
is rather likely that close to the short time interval $\delta t$ 
there will be another short time interval $\Delta t$, hence the denominator
in equation (10) will acquire a small value and the quantity 
$h(\delta t, \Delta t)$  will be comparably large.
Evidently, $h$ takes on values between 0 and 1, and we can either
consider the probability distribution $p(h)$, or the so-called 
''surviving function'' $P(h>H)$ that 
$h$ is larger than a certain value $H$. For a Poisson process,  with
time dependent rate parameters, it was shown that the latter probability 
is $P_{Poisson}(h>H)=1-H$ \cite{LEPRETI}.
The actual curve that results from paleomagnetic measurements was
shown in figure 2 \cite{CARBONE}, and is also given in figure 2
of the present paper.
Based on this, the second functional to be considered is
\begin{eqnarray}
F_{cluster}=\sum_{i=1}^{1000} (P(h>H_i)_{real}-{P(h>H_i)}_{num})^2  \; .
\end{eqnarray}
with $H_i=i \cdot 0.001$.

While the asymmetry of reversals is rather well understood in terms of
the field dynamics in the vicinity of an exceptional point of the 
spectrum,
the physical reason of the clustering property is less obvious.
In \cite{GAFD,LUCA} we had speculated that the 
appearance of clusters, which
can be visualized as a  ''devil's staircase'' of 
reversal events \cite{DEMICHELIS}, is a typical sign of 
''punctuated equilibrium'' \cite{GOULD} and self-organized criticality 
\cite{BAK}. In this respect it is also interesting to 
note the
tendency of the clustering property to increase with the degree 
of super-criticality \cite{LUCA}.

The third functional refers to the distribution function
of inter-reversal times (or residence times). The appearance of 
certain periods in the reversal sequence had been controversially 
discussed for a 
long time \cite{YAMAZ,LIU,WINKL}.
A breakthrough was achieved by Consolini and DeMichelis
who analyzed not the Fourier spectrum but the probability 
distribution 
of 
inter-reversal times \cite{CONSOLINI}. In this distribution 
they observed a clear sequence
of several maxima at multiples of approximately 95 kyr.
The physical reason for these maxima was identified as a stochastic resonance with 
the  Milankovic 
cycle of the 
Earth's orbit eccentricity \cite{LORITO}. 
The data from \cite{CONSOLINI} are shown again in figure 3, and the corresponding
functional is
\begin{eqnarray}
F_{rtd}=\sum_{i=1}^{90} (p(\tau_i)_{real}-{p(\tau_i)}_{num})^2
\end{eqnarray}
with $\tau_i=i\cdot1$ kyr.

Obviously, having defined the three functionals, there still remains 
an ambiguity in the choice of the relative weights for them.
We will test some reasonable
choices of the a-priori errors $\sigma_{shape}$, $\sigma_{cluster}$, and
$\sigma_{rtd}$
in the total functional
\begin{eqnarray}
F_{total}={\sigma_{shape}^{-2}} F_{shape}+ {\sigma_{cluster}^{-2}} F_{cluster} 
+{\sigma_{rtd}^{-2}} F_{rtd}
\end{eqnarray}
and check afterwards the correspondence of the resulting curves
with the paleomagnetic ones.

The very inversion is carried out by using a standard downhill
simplex method taken from ''Numerical Recipes'' \cite{RECIPES}.
Since we have  5 parameters to be determined, we use a simplex with
6 points. For each of those points in parameter space we solve the
forward problem for 20000 diffusion times which typically gives 
a few thousand reversals. From these reversals the functional
(16) is computed, and based on this evaluation the usual steps
of the downhill simplex method are performed, until an appropriate
stopping criterion is reached. While an individual run takes a few 
hours on a normal PC, the solution of the inverse problem
takes about one week.

\subsection{Results}

In the following we will 
present three solutions which result from choosing
different weights of the
three individuals functionals in (16).

Table 1 shows the obtained parameters of the three versions.
The second raw gives the parameter $C$, the third raw the 
shift parameter $\delta$. In the fourth raw the
critical value of the dynamo for the $\alpha(r)$-profile
with 
$\alpha_0=1+\delta$, $\alpha_1=
\alpha_3=0$, $\alpha_2=-6$ and $\alpha_4=5$ 
(see Eqs. 5-7) is shown, which differs 
from version to version due to the variation of $\delta$. The fifth 
raw gives the degree of supercriticality, i.e. $C/C_{crit}$, where
$C_{\rm{crit}}$ prescribes the amplitude of the 
critical $\alpha$ at which
the onset of dynamo action occurs. 
The raws 6, 7 and 8 show the 
noise level $D$, the (effective) diffusion 
time $T_d$, and the strength of the periodic 
forcing $\epsilon$, while  the 
last raw gives the resulting number 
of reversals per Myr for each version.

\begin{table}
\caption{Parameters of the three versions resulting from 
the downhill simplex inversion.}
\begin{indented}
\item[]\begin{tabular}{rrrrrrrrr}
\br
&${{C}}$&$\delta$&$C_{crit}$&$C/C_{crit}$&$D$&${{T_{d}}}$& {$\epsilon$} &$\#$/Myr\\
\mr
Version 1&57.0&-0.03&13.0&4.4&6.3&86.8 kyr&0.133  &2.05\\
Version 2&96.8&-0.16&8.99&10.8&8.0&63.7 kyr&0.123 &3.07\\
Version 3&147.8&-0.18&8.73&16.9&7.8&55.1 kyr&0.094&2.91\\
\br
\end{tabular}
\end{indented}
\end{table}

Figures 1, 2 and 3 show the curves for the
temporal dependence of the VADM, for the probability 
$P(h>H)$, and
for the RTD between reversals,
compared with the corresponding curves 
from the paleomagnetic data.

\begin{figure}[ht]
\begin{flushright}
\includegraphics[width=13cm]{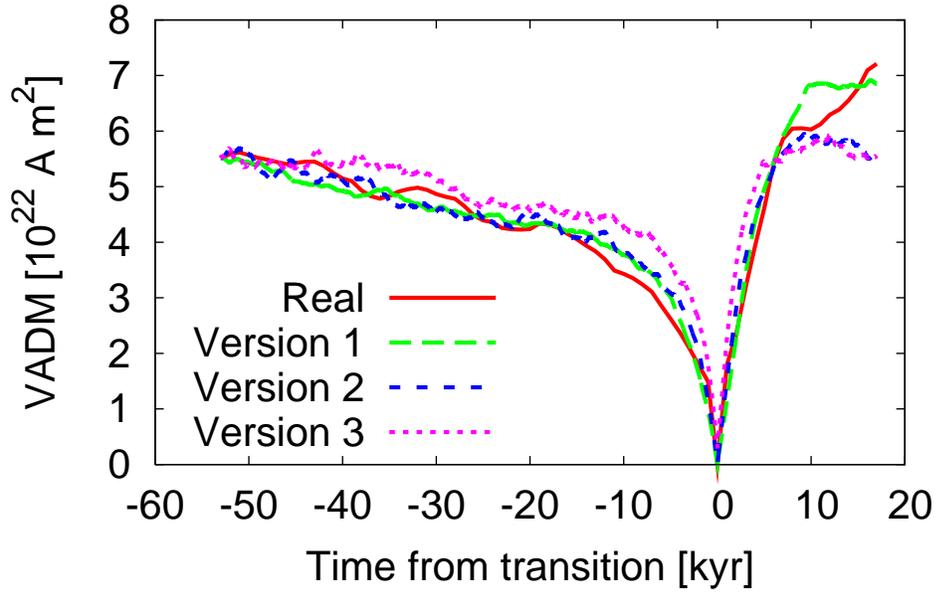}
\end{flushright}
\caption{Temporal evolution of the virtual axial dipole moment 
(VADM) during reversals. Comparison of the average over the last five 
real reversal data (taken from \cite{VALET}) 
with the average over 100 reversals for the
three versions resulting from different
solutions of the inverse problem.}
\end{figure}

\begin{figure}[ht]
\begin{flushright}
\includegraphics[width=13cm]{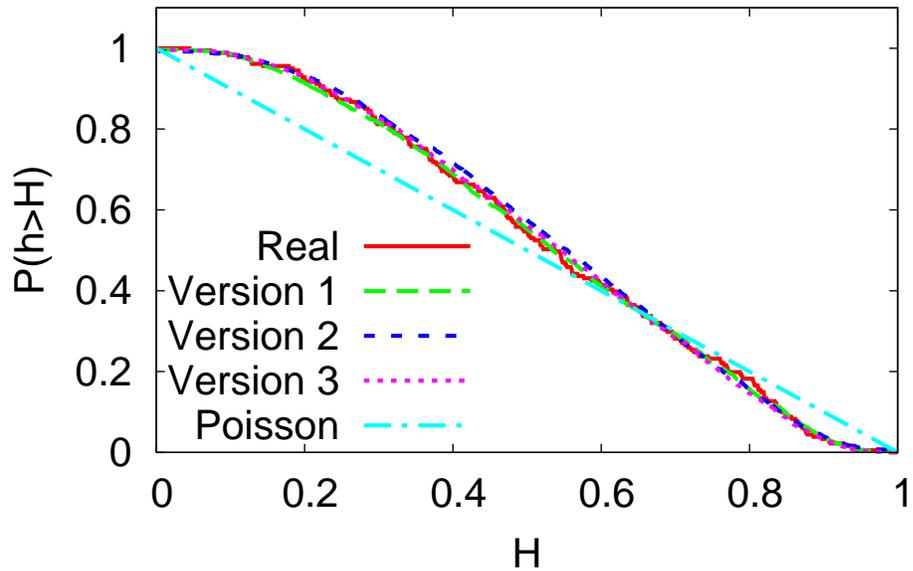}
\end{flushright}
\caption{Surviving function $P(h>H)$ for the real data 
(adopted from \cite{CARBONE,CK}) and
for the three versions.
The deviation from the straight line which would 
appear for a Poisson process 
indicates a significant clustering.}
\end{figure}

\begin{figure}[ht]
\begin{flushright}
\includegraphics[width=13cm]{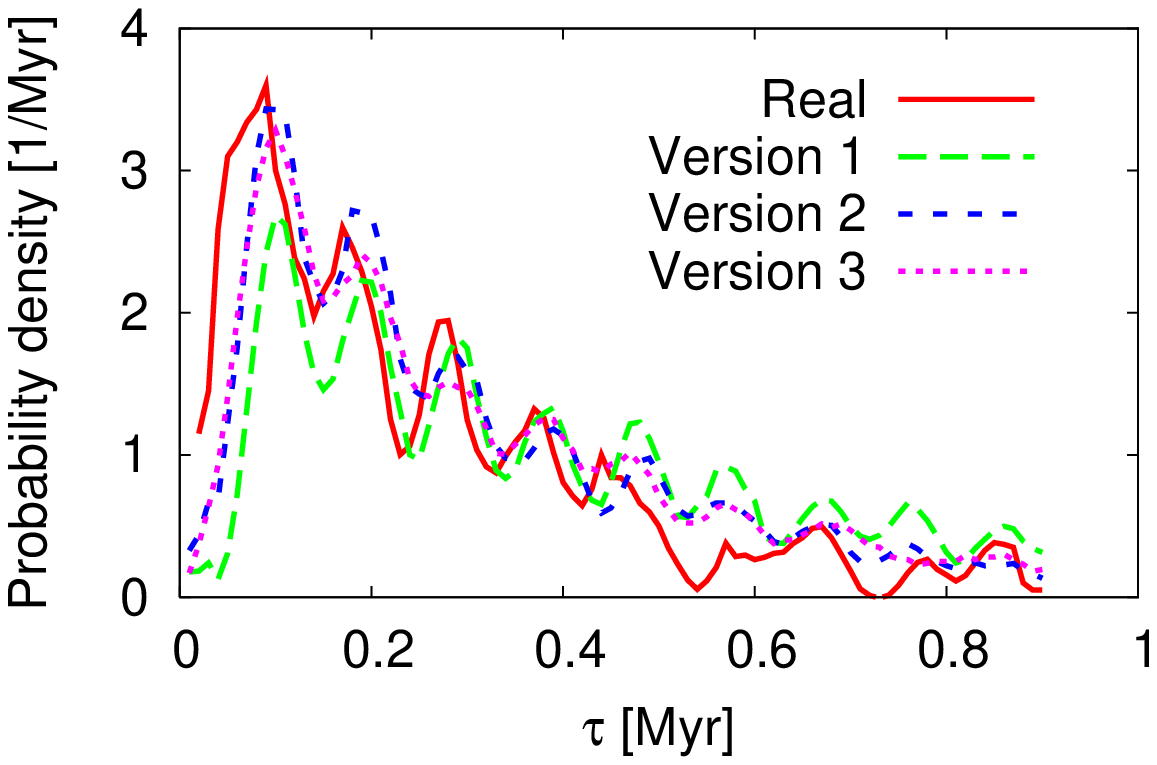}
\end{flushright}
\caption{Residence time distribution (RTD) for the real data 
(taken from 
\cite{CONSOLINI}) and 
for the three versions.}
\end{figure}

Obviously, version 2 gives the best correspondence with all three
reversal characteristics. Compared to this ''optimal'' version 2, 
in version 1 $\sigma_{rtd}$ was increased by a factor 2
which results in a deteriorated correspondence with the real
RTD curve in figure 3. In version 3, $\sigma_{shape}$
was increased by a factor 2 (compared to version 2), which leads 
to a deteriorated correspondence with the real VADM curve in
figure 1. Evidently, the large value of $C$ and the 
small $T_d$ provide reversals
that are faster than the really observed ones.

\section{Conclusions}
With version 2, we have obtained the best reproduction 
of the paleomagnetic 
input data for a 10 times supercritical dynamo, 
a relative strength of the periodic forcing of some 15 per 
cent, and an 
effective magnetic
diffusion time of approximately 65 kyr which is by a factor 3.5
smaller than the value that would result from the molecular 
conductivity.
The latter is perhaps the most important result of the inversion,
for the following reason: 
The conductivity enhancing effect of turbulence is a subject of ongoing
debate.
Only very few estimations exists for the turbulent diffusivities within the
Earth's outer core. 
However, their values are of extraordinary importance for the key parameters
of the geodynamo (e.g. Ekman number or the magnetic Prandtl number) as well as
for the estimation of turbulent transport properties and the destructive
influence of turbulence on magnetic field generation through turbulent field diffusion. 
The molecular values of the dimensionless parameters are
extremely small and cannot be realized in simulations so that usually enhanced
values are applied that shall resemble the effective (i.e. turbulent) quantities.
A more detailed information on the turbulent diffusivities therefore delivers
important information on the significance of the parameter space accessible in
global MHD simulations for the geodynamo.  
Furthermore, reality is more complicated, as it is very likely that 
a conductivity reduction due to turbulence would be anisotropic, 
because the small scale turbulence in a fast rotating object like the Earth
is subject to preferred directions parallel to the rotation axis (and also
along the dominating field component) so that the convection cells (that
determine the turbulent transport of physical quantities) are
oriented parallel to the rotation axis and elongated along the magnetic field.
It is well known that
an anisotropic conductivity 
could
have a tremendous effect on the selection between equatorial and axial
dipole solution \cite{TILGNER,ELSTNER}. Roughly speaking, axially aligned 
rotating columns
tend to decrease the conductivity for horizontal currents 
less  than the conductivity for vertical currents, and this effect 
leads to
a preference of the axial dipole compared to the equatorial dipole.
This could be an extremely important point since the conclusions 
of many geodynamo related papers 
rely  on an isotropic conductivity when looking for 
criteria for the selection of axial or equatorial dipoles.

The stunning agreement of paleomagnetic and 
numerical
reversal characteristics as shown in figures 1, 2 and 3 
gives support to our hypothesis that 
reversals are indeed noise
triggered relaxation oscillations in the vicinity 
of an exceptional
point of the spectrum of the dynamo operator.
In this respect it is important to note that
in particular
the time asymmetry
and the clustering property are intrinsic and robust 
properties
of the model that appear for very
wide regions of parameter (if not for all).
By solving the inverse problem we have 
not ''produced'' them, but 
have only fine-tuned the
dynamo parameters to fit optimally the paleomagnetic data.

We have carefully tried not to over-interpret our simple model
by  focusing  only on those parameters to be determined, and those 
functionals
to be minimized, that refer to the  temporal properties of 
reversal sequences, and not to any spatial features. 
This makes us optimistic that the results will prove robust when 
inversions of this kind will later be
repeated using more realistic dynamo models. Given that 
one downhill simplex run for our simple model takes already 
one week, one can 
imagine that corresponding runs with better dynamo models 
will lead to significant computational costs.

\ack
This work was supported by Deutsche Forschungsgemeinschaft
in frame of SFB 609.

\end{document}